# RESONANT INFILTRATION OF AN OPAL: REFLECTION LINESHAPE AND CONTRIBUTION FROM IN-DEPTH REGIONS

**Isabelle MAURIN**[1,2] **and Daniel BLOCH**[2,1]

(1) Laboratoire de Physique des Lasers, Université Paris 13, Sorbonne Paris-Cité

(2) CNRS, UMR 7538, 99 Avenue J-.B. Clément, F-93430 Villetaneuse, France

isabelle.maurin@univ-paris13.fr

daniel.bloch@univ-paris13.fr

*We analyze the resonant variation of the optical reflection on an infiltrated artificial opal made of transparent nanospheres. The resonant infiltration is considered as a perturbation in the frame of a previously described one-dimensional model based upon a stratified effective index. We show that for a thin slice of resonant medium, the resonant response oscillates with the position of this slice. We derive that for adequate conditions of incidence angle, this spatially oscillating behavior matches the geometrical periodicity of the opal, and hence the related density of resonant infiltration. Close to these matching conditions, the resonant response of the global infiltration varies sharply in amplitude and shape with the incidence angle and polarization. The corresponding resonant reflection originates from a rather deep infiltration, up to several wavelengths or layers of spheres. Finally, we discuss the relationship between the present predictions and our previous observations on an opal infiltrated with a resonant vapor.*

# I. Introduction

Photonic crystals offer the possibility to modify the distribution of optical modes, hence changing the properties of embedded absorbers or of resonant emitters. Conversely, a distribution of resonant particles inside the photonic crystal [1] may allow a fine tuning of the optical properties of the material, or can be of interest for a variety of sensors. An artificial opal, which is a (quasi-) crystalline arrangement of identical transparent nanospheres like glass nanospheres, is an approximate realization of a photonic crystal [2], obtained through soft chemistry. The opal voids easily allow for an infiltration by a resonant fluid (liquid, or gas). When an opal is prepared layer-by-layer in a Langmuir-Blodgett (LB) technique [3], the number of layers of nanospheres is kept under control, but a random hexagonal closed packed (*r.h.c.p.*) structure is generated, without a genuine crystalline order along the normal to the surface. Also, the optical properties of a photonic crystal, often analyzed through a reflection detection, strongly depend of the interface between the crystal itself and the external medium where detection is operated [4].

Here, we extend a calculation previously developed for the optics of an opal [4], to estimate on reflection the resonant effects of an infiltration. This is a type of "selective reflection" spectroscopy for a resonant medium embedded in the voids of the opal, where the optical reflection is monitored when the irradiation frequency is tuned around the resonance. In [4], the calculation, which applies to a broad range of wavelengths ($\lambda$) and sphere diameters (*D*), uses a one-dimensional stratified effective index defined through a height-dependent mix of the index of the glass spheres and of the voids. It has revealed convenient to emphasize the role of the first half-layer of spheres - "gap" region between the substrate and the compact opal - or of the last half layer, notably in reflection, as the "effective index" of this "gap" region is smaller than in the regions where the spheres are compactly arranged. The calculation is simple in its principle because it replaces the effect of a three-dimensional scattering by an *ad hoc* one-dimensional extinction coefficient. Nevertheless, it describes the reflection signal as interferences between the reflection from stratified layers, so that one recovers various standard predictions, notably Fabry-Perot type oscillations between the peripheral regions of the opal, and Bragg reflection peaks for well-chosen sets of wavelength $\lambda$ and incidence angle $\theta$. Technically, the continuous variations of the effective index are discretized to make the calculations tractable.

The stratified effective index model [4], relies on a linear formalism through a product of propagation matrices. This notably allows introducing a defect layer (see section III-4 of [4]). In the present work, the infiltrated material is assumed to exhibit simple resonances, as for a homogeneously-broadened gas (neglecting all effects of an internal motion) or a (weakly) absorbing liquid, and is considered as a perturbation in the effective index description of the opal [5]. In section II, we consider the effect of adding a single thin "slice" of a resonant material, which is a kind of a "defect layer". We show that, for a given location of the slice, the resonant change in reflection can single out the absorptive part of the resonant response, or the dispersive one, and generally yield a mix of these features. For simplicity, we always neglect the smooth wavelength dependence of the optical response of the bare opal itself [4]. Indeed, we always assume the resonance of the material to be sharper than the wavelength dependence of the opal, even around a Bragg reflection peak of the opal.

In a second step (section III), the resonant material is considered to be spatially distributed in the voids (interstices) of the opal, and we calculate the resonant effect by the coherent addition of the effect of the various of resonant slices, assumed to be weighted by the density of the voids. We hence show (section IV) that for a matching condition between the optical periodicity of the propagating wave (depending on the incidence angle) and the geometrical distribution of the opal, one can be sensitive in reflection to the contribution of relatively deep resonant layers. In section V, we finally compare these predictions of quick evolutions of the lineshape with experimental situations previously encountered with an infiltration of a Doppler-broadened gas of atoms undergoing thermal motion [6,7].

**II. Effect of a single slice of a resonant material**

The layered effective index model [4] deals with the optics of the opal assuming that the opal is stratified parallel to the substrate, as governed by a (discretized) index $N_{eff}(z)$. An extinction coefficient $\alpha(z)$ is introduced in $N_{eff}(z)$, despite the transparency of the opal spheres, in order to take into account the scattering, so that:

$$N_{eff}(z) = n_{eff}(z) - j\,\alpha(z)\lambda/4 = \Re e[N_{eff}(z)] + j\,\Im m[N_{eff}(z)] \qquad (1)$$

with $n_{eff}(z)$ a real index connected to the spatial variation -integrated over a plane parallel to the substrate- of the opal density (see fig.1a); for simplicity, $\alpha(z)$ is assumed to be constant with $z$ (see discussion in [4]), and independent of the propagation direction of the beam. Here, because we assume the presence of a slice of resonant infiltrated material, of a thickness $\Delta z$, located between $z_0$ and $z_0 + \Delta z$, we introduce an extra resonant (complex) contribution to the local optical index:

$$\Delta N(\omega) = \Re e[\Delta N(\omega)] + j\,\Im m[\Delta N(\omega)] \qquad (2)$$

which has to be added in eq. 1 (for $z_0 \leq z \leq z_0 + \Delta z$). The value of $\Delta N(\omega)$ depends on the (nearly resonant) excitation frequency $\omega$. In eq.(2), $\Re e[\Delta N(\omega)]$ corresponds to the dispersive part of the resonance, and $\Im m[\Delta N(\omega)]$ to the absorption of the infiltrated material (assuming $\Im m[\Delta N(\omega)] < 0$).

To calculate the effect of the resonant infiltrated slice, a perturbation expansion is applied to the calculation of the reflection coefficient [8]. Note that in a perturbative approach, the reflection coefficient in amplitude, and in intensity, behave similarly. Hence, the resonant change in reflectivity solely depends on the product $\Delta N(\omega).\Delta z$ because of linearity. Moreover, this linearity applies separately to $\Re e[\Delta N(\omega)].\Delta z$, and to $\Im m[\Delta N(\omega)].\Delta z$, so that the frequency lineshape of the reflective response of this resonant slice is simply deduced from the knowledge of the complex resonance $\Delta N(\omega)$.

Figure 2 shows a variety of situations, where calculations are for realistic parameters (corresponding for $D$ and $\lambda$ values to experimental situations that we have analyzed [6,7], and to a realistic extinction coefficient, see [4]). As a general result, the predicted change $\Delta R(z_0,\omega)$ in the reflection coefficient $R$ oscillates with the position $z_0$ of the resonant slice. The spatial period of this oscillation is determined by the optical propagation, showing a dependence on the wavelength of the excitation $\lambda$ (fig.2b) and on the incidence angle (fig.2c). Naturally, this spatial period depends not only on the vacuum wavelength $\lambda$, but also depends on the (averaged) value of $n_{eff}(z)$, defining the "effective" wavelength in the opal. The oscillation with the position of the resonant slice embedded in a layered material generalizes a well-known behavior for the reflective contribution of an additional slice in a homogeneous medium (see *e.g.* [9]). In addition to this oscillating behavior, the response is spatially damped, as a consequence of the phenomenological attenuation coefficient $\alpha(z)$ introduced to take into account the effect of light scattering.

A closer look on this oscillation reveals a (spatial) phase shift when turning the resonance from dispersive to absorptive (fig. 2d). This phase-shift can be traced back to the well-known fact that, relatively to the incident driving field, the resonant (field) response of the driven material itself exhibits a phase-shift when the response goes from dispersion to absorption. An analogous but weaker phase-shift is also found when switching the incident polarization from one to the other principal polarization, *i.e.* from TE to TM (see fig. 2e). This phase-shift associated to polarization would not be predicted for a resonant slice in a homogeneous isotropic medium: it is a genuine signature of the stratified medium in which the resonant layer is embedded [4, 6-7]. It is indeed the propagation to this resonant layer, and from it, which differs for a TE- or TM- polarized beam, owing to a polarization-dependent combination of transmission and reflection coefficients through the successive layers. This justifies that the polarization dependence of these oscillations is small for weak incidence angles, and becomes possibly sizeable for large incidence angles, notably in the vicinity of the opal "Brewster" angle (see [4]). This kind of birefringence can be viewed as the result of the anisotropy introduced by the organization of the rather thick opal parallel to the window.

**III. Spatially distributed infiltration, matching and deep layers contribution**

The linearity of our calculation also allows evaluating the global effect of a resonant infiltration by summing the contribution of elementary slices. Only the density of the resonant material must remain weak enough to justify a perturbation expansion on the reflection coefficient. The change of reflectivity becomes:

$$\Delta R(\omega) = \int_0^\infty \frac{\Delta R(z,\omega)}{\Delta z} dz \qquad (3)$$

with $\Delta R(z, \omega)$ the change of reflectivity for an elementary slice, located in $z$ and of a $\Delta z$ thickness, as calculated in section II.

At this step, the spatial distribution of the infiltrated material must be considered. Indeed, if one simply assumes a constant density for the infiltrated material, the oscillations of the response $\Delta R(z, \omega)$ for an elementary slice (section II, and fig.2) just lead to a coherence length limited to the (reduced) period of the oscillation. Above a ~1 rd shift, distant slices no longer interfere constructively. Practically, it is more natural to assume that the resonant material fills up only the voids of the opals, so that its density must follow the [1 - $f(z)$] spatial distribution (see fig 1.b). This density, which also governs the corresponding (sliced) optical index $\Delta N(\omega)$, is periodical after the first half layer of spheres.

A match between the optical periodicity of the response (for an elementary slice) and the geometrical periodicity of the infiltration occurs for specific sets of wavelength and incidence angle. Interesting situations then appear, as can be seen with a plot of the partial integration:

$$\Delta R_z(\omega) = \int_0^z \frac{\Delta R(z,\omega)}{\Delta z} dz \quad (4)$$

In fig. 3, this partial integration from the interface is shown for a variety of incidence angles, with $\Delta N(\omega)$ proportional to the $[1 - f(z)]$ distribution, and purely real (fig. 3a) or purely imaginary (fig. 3b). These partial integrations always exhibit a fast oscillating behavior with the boundary of the integration. This had to be expected from fig. 2. Interestingly, and despite these oscillations, the partial integration increases with the distance over several optical wavelengths for some incidences [notably in fig.3a for $\theta = 50°$ and dispersive response $\Delta N(\omega) = \Re[\Delta N(\omega)]$, or for $\theta = 52°$ and absorptive response $\Delta N(\omega) = j \Im[\Delta N(\omega)]$ in fig.3b]. This is like if the coherence length would be now governed by a kind of "supergrating", involving the geometrical periodicity of the material density and the optical periodicity of the response. The already-mentioned phase shift between dispersive-type and absorptive-type resonance justifies that the most sizeable responses for each kind of resonance are not obtained for the same incidence. Analogous changes of the optimal incidence angle to get these long "coherence lengths" can also be found when comparing TE and TM polarizations, as a result of the phase shift between TE and TM elementary responses. The essence of this regime of "long coherence length" is that a notable contribution to the resonant signal originates in deep layers of the opal.

Additionally, one notes that for the optimal incidence (*e.g.* $\theta = 52°$ for absorptive response), the partial integral sticks to a long-distance asymptotic value (fig.3b), instead of growing continuously when increasing the domain of integration. This is an effect of the attenuation coefficient. Conversely, even for a small mismatching (*e.g.* $\theta = 50°$ for absorptive response), the overall integration exhibits a small decrease for long distances after having reached an extreme. In some cases, like for 40° or 52° (dispersive response), the integral apparently sticks to a given value, already determined by the first layer. Indeed, the first half-layer, mostly made of voids and allowing a high density of the infiltrated material, can have a predominant influence. This behavior can appear as an offset, which may be discriminated in some cases (see *e.g.* the case of gas of "moving" atoms [6]). In the same manner, the [1 - *f(z)*] distribution includes a residual non-modulated infiltration (for a perfect opal, this floor represents a ~ 9% density): its effect is equivalent to a tiny (resonant) change of the sphere index.

**IV. Resonant lineshapes**

The linearity of the treatment, applied in the previous section separately to a purely real or purely imaginary resonant change $\Delta N(\omega)$, is the basis to predict, from the integrated value $\Delta R(\omega)$ (eq. 3, which is the long distance limit of the partial integration discussed in section III and fig.3), the reflection lineshape for the infiltrated opal. The marked sensitivity of the amplitude $\Delta R(\omega)$ to the incidence angle is illustrated in fig.4, where $\Delta R(\omega)$ is plotted as a function of $\theta$, when the resonant response is either purely dispersive ($\Im m\,[\Delta N(\omega)] = 0$), or absorptive ($\Re e\,[\Delta N(\omega)] = 0$). The sensitivity to the resonant medium is the highest around the angle satisfying a matching between the opal periodicity and the optical propagation, but marked differences can be seen between absorptive and dispersive responses. This implies that for a given lineshape of the intrinsic optical resonance of the infiltrated material, the reflection signal exhibits considerable variations of the lineshape under a small change of the incidence angle. Similar strong variations arise when changing polarization from TE to TM.

These quick variations are illustrated in fig.5, where for simplicity we assume for the infiltrated material a (complex) Lorentzian response:

$$\Delta N(\omega) \propto - [1 - f(z)]\,[(\omega - \omega_0) - j\gamma]^{-1} \qquad (5)$$

with $\omega_0$ the center of the resonance, and $\gamma$ the width of the Lorentzian response of the infiltrated material.

It is worth noting that, in order to compare the resonant changes in the reflection signal, we have plotted in fig.5 the *relative* change of reflectivity $\Delta R(\omega)/R$. In particular, for TM polarization (see inset of fig.5), the non resonant reflection is pretty weak, but exhibits notable variations : this is because we are in the region of a Brewster-like incidence [4]. In TE polarization, $R$ exhibits a maximum around 52°, which can be actually traced back to a second-order "Bragg peak", as shown to exist in [4]. The resonant variation predicted for the reflectivity lineshape $\Delta R(\omega)/R$ with the incidence angle is much larger and essentially different in shape, than the one governed by high-order Bragg reflection, whose shape may eventually become a dip through complex interferences in reflection which do not all satisfy the Bragg-matching condition. Anyhow, as long as the density of the resonant medium mimics the opal geometry, the coincidence is intrinsic between the incidence yielding spatial coherence of the response of the infiltrated material, and the one of Bragg reflection, as corresponding to a coherent summing of the successive reflection over the opal layers. This can be verified when calculations are performed for a different set of ($\lambda$, $D$) values.

**V Discussion and connection with possible experiments**

The above approach was initially developed to help interpreting experiments with an infiltration of a low-density atomic gas, sensitive to the thermal atomic motion, inside the interstitial regions of an LB opal [6]. In addition to the expected Doppler-broadened structures resulting from the thermal atomic motion, we had observed sub-Doppler contributions, but only for a range of oblique incidences, and with rapid variations of these narrow lineshapes when varying the incidence angle. The profound physical origin of this sub-Doppler contribution, probably connected to a velocity-selective transient behaviour in the confined interstitial responses, has remained elusive until now [6,7]. Various optical considerations had led us to attribute [6] the narrow signal to a resonant response of atoms three-dimensionally confined inside the opal. In spite of the physical differences between our present local index approach, applicable only for a motionless resonant medium, and the Doppler-broadened gas, it is interesting to note that the sets of lineshapes calculated in fig. 5 exhibit some similarities [10] with the narrow sub-Doppler structures appearing in [6]. This can appear as an additional indication that these narrow structures are truly related to an in-depth resonant response inside the opal. Our model notably predicts the possibility of dramatic changes in the lineshape for a given incidence, and for a modest change in the wavelength. This resembles our observations [6], when comparing results on the Cs doublet resonance lines, with the $D_1$ line at $\lambda_1 = 894$ nm, and the $D_2$ line at $\lambda_2 = 852$ nm. A sensitivity to polarization is here predicted, analogous to what was dramatically observed in experiments. Also, for a more radical change in the wavelength, like when turning to the second resonance line of Cs at $\lambda = 455$ nm, one predicts here a faster variation of the lineshape with the incidence angle, as has been recently observed [7].

Anyhow, the present modeling cannot have the ambition to deal with the Doppler effect associated to atomic thermal motion, and it is restricted to homogeneously broadened simple resonant fluid materials. Aside from the intrinsic limitations of our one-dimensional stratified medium, several limitations may apply to the quantitative findings of our model. First, the linearity argument applies in principle only for a low signal or density of resonant particles. Here, careful checks may be needed as the calculation for an opal is practically sliced in up to ~ $10^3$ slices (see *e.g.* the estimation for numerical calculation in [4]). Second, a glass opal prepared by a LB method tends to exhibit numerous structural defects, which increase with the distance to the surface as corresponding to successive steps of deposition. Such defects should impact the periodicity of the [1- $f(z)$] density function, and reduce the possibility to generate a coherent resonant response from the most remote layers. The observation of a non resonant high-order Bragg peak may provide an indication of the quality of the opal, sustaining the predicted sharp variations with the incidence. Note that, with polystyrene spheres, highly ordered LB opals can be produced. They could be compatible with various infiltrating materials, although not relevant for experiments with alkali metal vapors.

Finally, an important limitation of the model is that in its essence, the one-dimensional stratified medium ignores the light scattered by the opal, considered only through phenomenological losses. This scattered resonant light also contributes to the excitation of the infiltrated material, and the phase of the corresponding excitation (not necessarily random) usually differs from the one resulting from the excitation by the propagating field. As an alternate asymptotic regime, one may consider that the resonant excitation is only induced by a purely incoherent scattering. This leads again to a periodical spatial distribution of field emission by the infiltrated material, but with modified quantitative predictions. The general behavior, with its sharp dependence on the incidence angle and on polarization, would nevertheless remain similar, while the coincidence with a high-order Bragg peak should disappear. This says that even if it makes it difficult to assess the ordering quality of the opal by an analysis of the variations of the resonant signal, the general type of behavior that we describe is robust relatively to the specific hypotheses of the model.

**VI Conclusion**

The above simplified one-dimensional method has allowed considering the coupling of a resonant material with an opal, up to the possible predictions of lineshapes, a task which would be formidable for a purely three-dimensional numeric approach. By rather general arguments, we have shown that infiltration of relatively deep regions of a photonic crystal can be observed for a well-chosen narrow range of experimental considerations. Our numerical evaluations were performed in view of already performed experiments, in a regime far away from the main bandgap of the opal. Nevertheless, the enhanced sensitivity to the resonant material tends to occur close to a (high-order) Bragg condition, because the constructive interferences needed for a spatially coherent resonant excitation of the medium are driven by the same incident field allowing the interferences leading to the Bragg condition. This may naturally change if the exciting field is induced by the scattered light. For longer wavelengths (and unchanged sphere size), in the vicinity of the first-order Bragg reflection peak, our model could help understanding to which extent a nearly-forbidden propagation becomes partially allowed with the resonant infiltration [11]. Because the depth of the observed resonant response remains however confined at maximum to few wavelengths, our predictions of a sharp sensitivity to incidence angle and polarization should apply as well to a thick opal prepared by sedimentation. Finally, our conclusions can be straightforwardly extended to the situation of an inverse opal, or to the detection of resonant transmission in a sufficiently thin LB opal. An extension of this method to the spectroscopy of an infiltrated photonic fiber may even be considered on the basis of a scalar approach such as proposed in [12].

**Acknowledgements**

Work supported by the ANR project "Mesoscopic gas" 08-BLAN-0031. We thank the referee for drawing our attention to references 5 and 12.

**References**

[1] Yu. A. Vlasov, V. N. Astratov, O. Z. Karimov, A. A. Kaplyanskii, V. N. Bogomolov A. V. Prokofiev, *Existence of a photonic pseudogap for visible light in synthetic opals,* Phys Rev B **55**, R13357 (1997)

[2] C. López, *Materials aspects of photonic crystals*, Advanced materials, **15**, 1679-1704 (2003).

[3] S. Reculusa and S. Ravaine, *Synthesis of colloidal crystal of controllable thickness through the Langmuir-Blodgett technique,* Chemistry of materials, **23**, 598 (2003)

[4] I. Maurin, E. Moufarej, A. Laliotis, D. Bloch, *The optical interface of a photonic crystal: modelling an opal with a stratified effective index*, ArXiv v2; *Optics of an opal modeled with a stratified effective index and the effect of the interface*, submitted

[5] Such a perturbative approach has already been used, but not for a spectroscopic resonance, in a more general context of a layered medium, see : P. Bertrand, C. Hermann, G. Lampel, J. Peretti, and V. I. Safarov, *General analytical treatment of optics in layered structures: Application to magneto-optics*, Phys. Rev. B **64**, 235421 (2001)

[6] P. Ballin, E. Moufarej, I. Maurin, A. Laliotis, D. Bloch , *Three-dimensional confinement of vapor in nanostructures for sub-Doppler optical resolution*, Appl. Phys. Lett., **102**, 231115 (2013)

[7] E. Moufarej, I. Maurin, I. Zabkov, A. Laliotis, P. Ballin, V. Klimov, D. Bloch, "*Infiltrating a thin or single layer opal with an atomic vapour: sub-Doppler signals and crystal optics*", EPL **108**, 17008 (2014)

[8] In all the following -see discussion in section V-, we assume that the elementary resonant slice has a thickness $\Delta z$ much smaller than the sphere size and than the wavelength. Also, the resonant change of reflectivity is assumed to be small relatively to the non resonant eflectivity. Of course, this assumes that this non resonant reflectivity has not dropped down to

zero, as it may occur accidentally, *e.g.* in specific Brewster-like situations assuming the opal to be ideal. The non perturbative regime for an opal would extend the findings for a flat interface described in : A. M. Akulshin, V. L. Velichansky, A. I. Zherdev, A. S. Zibrov, V. I. Malakhova, V. V. Nikitin, V. A. Sautenkov, G. G. Kharisov, *Selective reflection from a glass–gas interface at high angles of incidence of light*, Sov. J. Quant. Electronics, **19**, 416 (1989).

[9] G. Nienhuis, F. Schuller, M. Ducloy, *Nonlinear selective reflection from an atomic vapor at arbitrary incidence angle*, Phys. Rev. A **38**, 5197(1988), see also the appendix of M. Chevrollier, M.Fichet, M. Oriá, G. Rahmat, D.Bloch, M. Ducloy, *High resolution selective reflection spectroscopy as a probe of long-range surface interaction : measurement of the surface van der Waals attraction exerted on excited Cs atoms,* J. Phys. II France **2**, 631 (1992).

[10] Note that in ref. 6, a frequency-modulation (FM) technique is used, which actually yields $\Delta R(\omega)/d\omega$.

[11] P. J. Harding, P.W.H. Pinkse, A. P. Mosk, W. L. Vos, *Nanophotonic hybridization of narrow atomic cesium resonances and photonic stop gaps of opaline nanostructures*, Phys. Rev. B **91**, 045123 (2015)

[12] Tanya M. Monro, D. J. Richardson, N. G. R. Broderick, P. J. Bennett, *Holey Optical Fibers: An Efficient Modal Model*, J. Lightwave Tech., **17**, 1093 (1999)

**Figure captions**

Fig. 1: (a) stratified layered effective index for an opal, including an additional perturbative slice $\Delta N.\Delta z$ located in $z_0$; (b) spatial distribution $[(1 - f(z)]$ of the voids, and of the infiltrated material.

Fig. 2: Relative change in the opal reflectivity as a function of the position $z_0$ of an elementary additional slice and of the external incidence angle $\theta$. (a) scheme of the set-up; (b) comparison between $\lambda_1 = 852$ nm and $\lambda_2 = 426$ nm, with $\theta = 45°$ and TE polarization; (c): comparison between $\theta = 20°$ and $\theta = 60°$, with $\lambda = 852$ nm and TE polarization; (d) comparison between a dispersive contribution (labelled Re) for $\Delta N(\omega).\Delta z$ real, and the equivalent absorptive one for $\Delta N(\omega).\Delta z$ imaginary (labelled Im), for $\lambda = 852$ nm, $\theta = 60°$, and TE polarization; (e) comparison between TE polarization and TM polarization, with $\lambda = 852$ nm and $\theta = 45°$ ($R = 4.7\%$ for TE, $R = 0.56\%$ for TM). The opal is assumed to be made of 1µm diameter glass spheres, with a glass index $n = 1.4$ identical to the index of the substrate. The extinction coefficient replacing the scattering is chosen to be $\alpha = 2.10^5$ m$^{-1}$, and the LB opal is made of 20 layers (or more) of glass spheres, in order to get a result independent of the opal thickness. One has taken $\Delta N(\omega).\Delta z = 10^{-11}$ nm for (b,c,e), and (d) when $\Delta N(\omega)$ is real, and $\Delta N(\omega).\Delta z = -j\, 10^{-11}$ nm for (d) when $\Delta N(\omega)$ is imaginary.

Fig. 3: Relative change in the reflectivity $\Delta R_z(\omega)/R$ as a function of the limit $z$ of the infiltrated region for the indicated incidence angles: (a) dispersive resonance $\Delta N(\omega) = 5.10^{-4}$; b) absorptive resonance $\Delta N(\omega) = -5.10^{-4}\, j$. The opal features are the same as in fig. 2, polarization is TE, $\lambda = 852$ nm.

Fig.4: Relative change in the reflectivity $\Delta R(\omega)/R$ for a wholly filled opal as a function of the incidence angle, for a dispersive resonance (labelled Re) $\Delta N(\omega) = 5.10^{-4}$, and for an equivalent absorptive resonance (labelled Im) $\Delta N(\omega) = -5.10^{-4} j$. The opal features are the same as in fig. 2, polarization is TE, $\lambda = 852$ nm.

Fig 5 : Calculated frequency lineshapes of the relative change in the reflection signal for an infiltration exhibiting a (complex) Lorentzian response, and for incidence angles as indicated. Polarization is: (a) TE; (b) TM. The scale of relative amplitude is the same for all curves, and one has respectively R = 5.93%; 10.32%; 7.33%; 6.32% for 48°, 52°, 54°, 56° in TE polarization, and = 0.45%; 0.29%; 0.14%; 0.08% in TM polarization. The inset shows the nonresonant behaviour of the reflection with the incidence angle, notably illustrating a second-order Bragg diffraction for TE, and a Brewster-like incidence in TM (same 2011wavelength and opal sphere diameter as in fig.2)

FIGURE 1

(a)

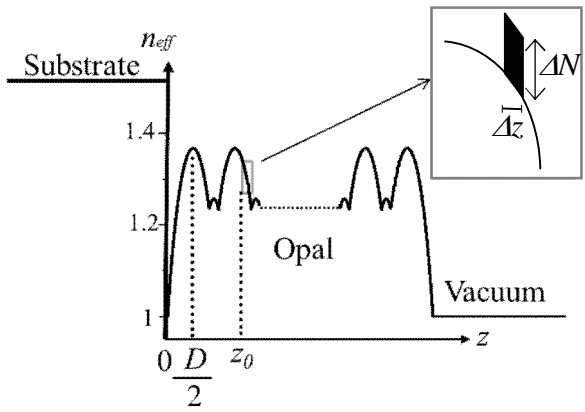

(b)

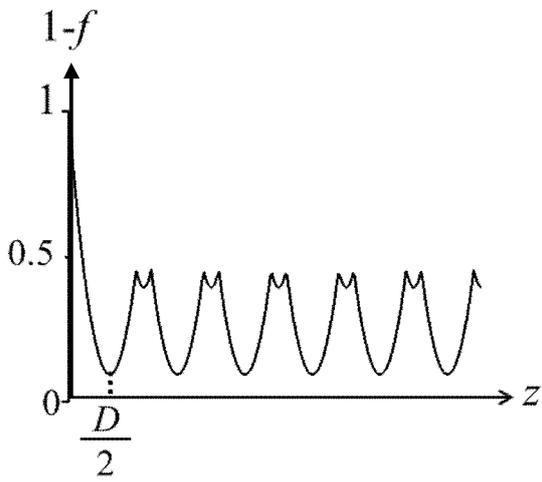

FIGURE 2

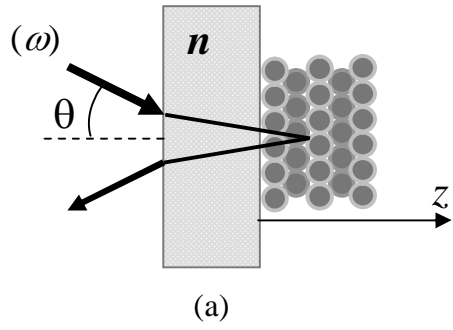

(a)

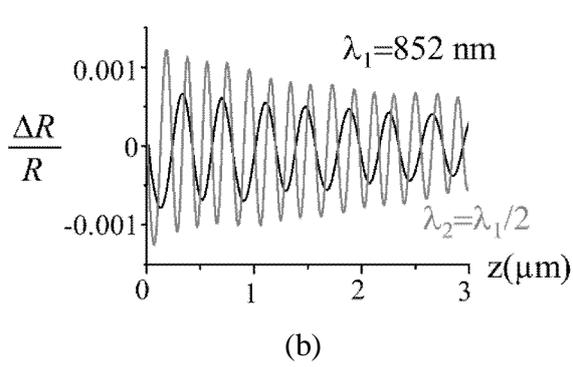

(b)

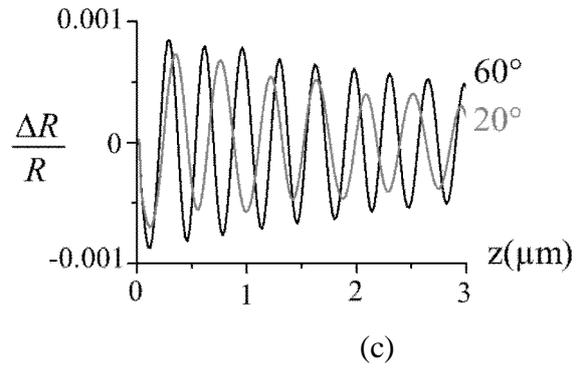

(c)

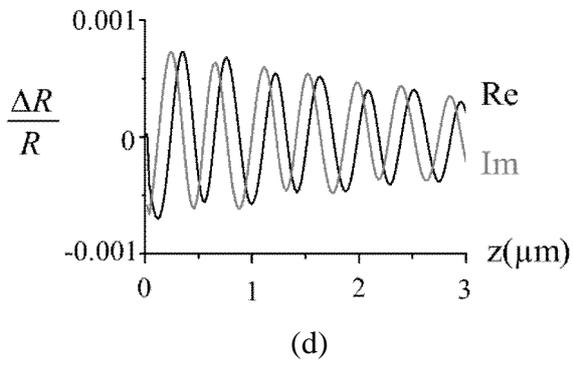

(d)

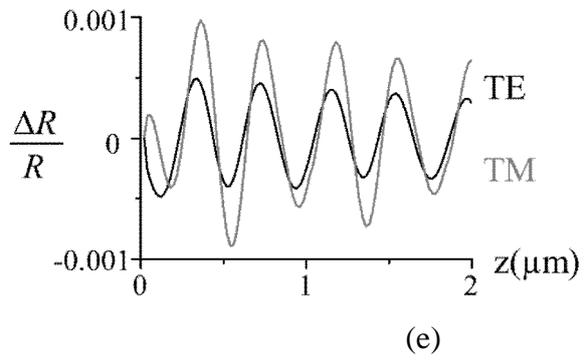

(e)

FIGURE 3

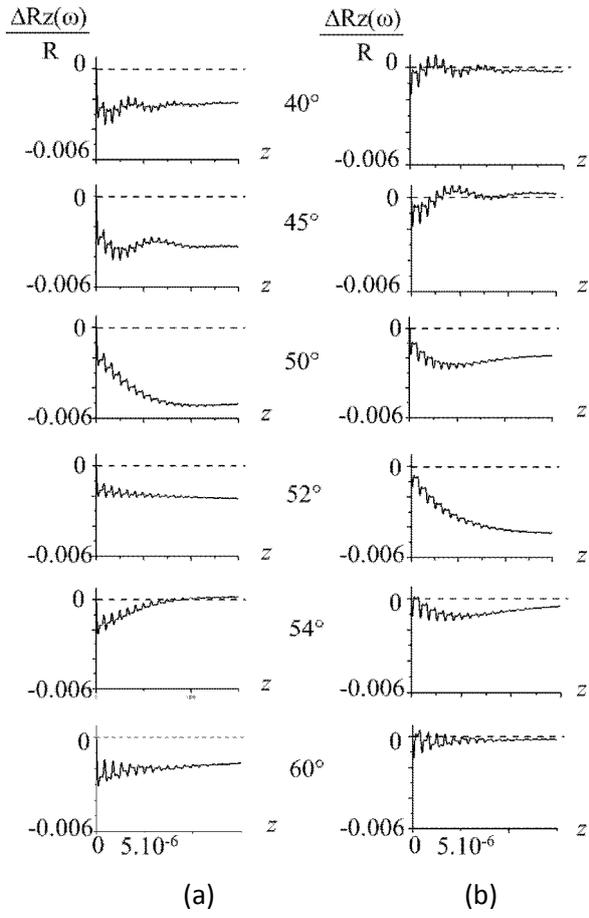

FIGURE 4

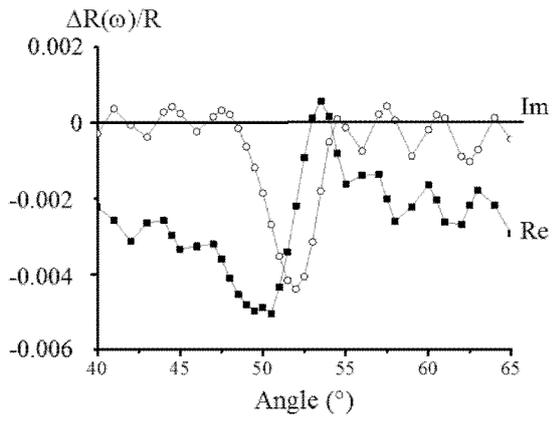

FIGURE 5

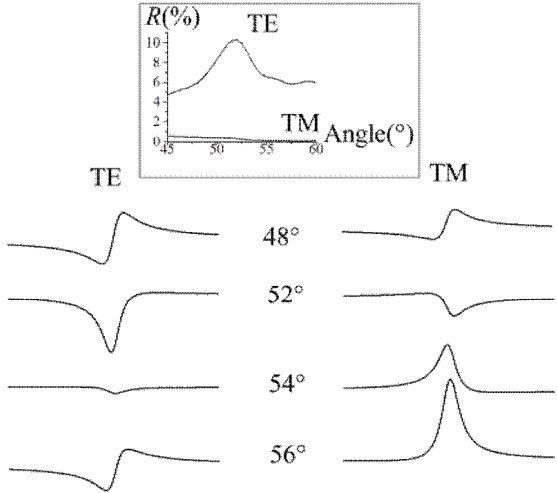